# Single-molecule Electroluminescence and Beyond

YAO ZHANG, YANG ZHANG AND ZHENCHAO DONG
HEFEI NATIONAL LABORATORY FOR PHYSICAL SCIENCES AT THE MICROSCALE AND SYNERGETIC INNOVATION CENTER
OF QUANTUM INFORMATION & QUANTUM PHYSICS, UNIVERSITY OF SCIENCE AND TECHNOLOGY OF CHINA

## ABSTRACT

A scanning tunneling microscope (STM) can do more than atomic imaging and manipulation. Its tunneling current can also be used for the excitation of light, converting electron energy to photon energy. STM based single-molecule electroluminescence can be realized by adopting a combined strategy of both efficient electronic decoupling and nanocavity plasmonic enhancement. The emission intensity, upon optimized material combination for the molecule, spacer, tip, and substrate, can be strong and stable enough for performing second-order photon correlation measurements. The observation of an evident photon antibunching effect demonstrates clearly the nature of single-photon emission for single-molecule electroluminescence. Strikingly, the spectral peak of a monomer is found to split when a molecular dimer is artificially constructed through STM manipulation, which suggests that the excitation energy from tunneling electrons is likely to rapidly delocalize over the whole molecular dimer. The spatial distribution of the excitonic coupling for different energy states in a dimer can be visualized in real space through sub-nanometer resolved electroluminescence imaging technique, which correlates very well with the local optical responses predicted in terms of coherent intermolecular dipole-dipole coupling. Furthermore, a single molecule can also couple coherently with a plasmonic nanocavity, resulting in the occurrence of interference-induced Fano resonance. These findings open up new avenues to fabricate electrically driven quantum light sources as well as to study intermolecular energy transfer, field-matter interaction, and molecular optoelectronics, all at the single-molecule level.

## INTRODUCTION

The conversion of electrical energy to photon energy is an important scientific issue. Research on single-molecule electroluminescence (SMEL), apart from its potential as a quantum light source, can provide deep insights into how a single molecule emits photons upon electrical excitation, particularly on how the spatial distribution of electronic states within a molecule as well as the local molecular arrangement and environment can affect the light emission behavior. Thanks to the highly localized nature of tunneling electron excitations in a scanning tunneling microscope (STM), STM induced luminescence (STML) is by far the most suitable technique for the SMEL studies down to the atomic scale [1-5]. The capability of STML to acquire simultaneously both topological and photon images in addition to spectral information makes it a powerful tool to study the photon emission properties of a single molecule comprehensively in both spatial and energy domains. Usually, the light emission of a single molecule would be quenched if it is directly adsorbed on the metal surface because of direct electron transfer and resultant fast non-radiative decay processes [6]. To suppress the fluorescence quenching, a decoupling layer such as a molecular multilayer [7, 8], a thin insulating layer [1-4, 9-11], or a built-in spacer [12, 13] must be introduced in order to obtain the molecule-specific luminescence arising from intramolecular optical transitions. A well-decoupled single molecule would behave as a two-level quantum system and can thus serve as a single-photon emitter. On the other hand, the photon signal of a single molecule near a metal surface is believed to be very weak. Fortunately, upon appropriate material selection (e.g., coinage metals) and tip fabrication, the metallic tip and substrate in a STM could de-fine a plasmonic nanocavity, through which the radiation from a single molecule could be dramatically enhanced because of the presence of a strong and localized electromagnetic field [7, 14, 15]. The introduced nanocavity plasmon (NCP) plays not only an important role in enhancing the molecular emission but also can interact with a single molecule, which can result in changes in the spectral features of electroluminescence [10, 16-18]. Furthermore, if there are interactions between single molecules adsorbed on the substrate, the spectral features of STML may be different from that of an isolated single molecule and the resultant information on spectral changes can then be used to unravel the type of interactions between the molecules [4, 19, 20].

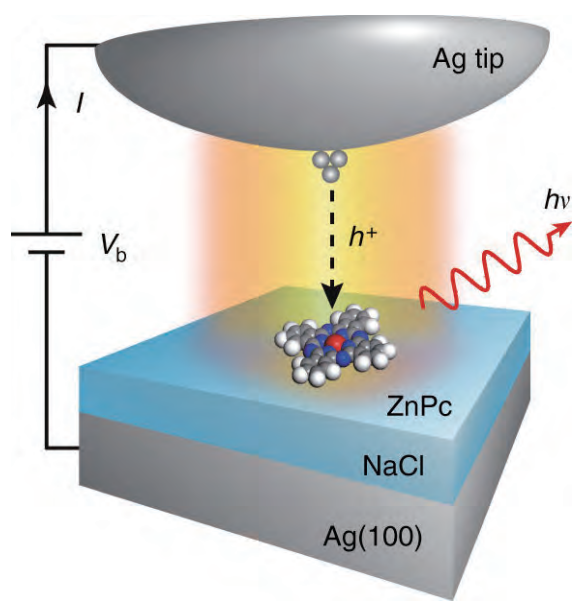

**Fig. 1:** Schematic of STM induced fluorescence from a single ZnPc molecule. Reprinted with permission from Macmillan Publishers Ltd: [Nature Communications], (L. Zhang, et al., Nature Communications, 8, 580 (2017)) [9].

In this article, we shall take the zinc-phthalocyanine (ZnPc) molecule as an example to summarize our recent works on single-molecule electroluminescence and related progresses. By adopting a combined strategy of electronic decoupling and NCP enhancement, we achieve stable and intense SMEL from an isolated neutral ZnPc molecule and further demonstrate its single-photon emission behavior through the observation of clear photon antibunching effects [9]. By further pushing the isolated single ZnPc molecules together to form a dimer in contact, the spectral features are found to change dramatically, which can be attributed to the coherent intermolecular dipole-dipole interaction [4].

However, if tunneling electrons initially excite the NCP alone rather than the molecule directly, the coherent interaction between the single molecule and the NCP will result in quantum interference and the appearance of Fano dips in the spectral features [10]. All these findings demonstrate that single-molecule electroluminescence can provide new opportunities to explore the emission properties of single molecules and the coupled systems at the single-photon and sub-nanometer level, which can help in the development of molecule-based quantum light sources, light-harvesting structures and nano-optoelectronics.

## ELECTRICALLY DRIVEN SINGLE-MOLECULE SINGLE-PHOTON EMISSION

Single-photon sources play important roles in quantum information and quantum computing technologies. Electrically driven single-molecule light emitters are considered to be promising candidates for single-photon sources. Although electrically driven single-photon emitters have been investigated in organic light-emitting diode devices by using dispersed single molecules, it is difficult to eliminate the influences from environment. The STML technique provides the capability to locally excite molecular electroluminescence by highly-confined tunneling electrons, and also the ability to enhance the emission signals by the local fields of the NCP. Efficient electronic decoupling and nanocavity plasmonic enhancement are two issues that are crucial for the achievement of STM based single-molecule electroluminescence. Figure 1 shows the schematic diagram to achieve single-molecule electroluminescence from an isolated ZnPc molecule that is electronically decoupled by a thin sodium chloride (NaCl) layer on a Ag (100) surface. The appropriate thickness of the NaCl decoupling layer is crucial for suppressing fluorescence quenching and generating the molecule-specific electroluminescence. The molecular fluorescence is excited by the localized tunneling electrons and further enhanced by the plasmonic nanocavity composed by the silver tip-substrate structure. Figure 2(a) shows the STML spectra of an isolated single ZnPc molecule adsorbed on 3 and 4 monolayers (MLs) of NaCl (as blue and red curves), respectively. The occurrence of sharp molecule-specific emission peak at ∼1.9 eV is a clear indication for the good performance of NaCl as a decoupling layer and for the realization of single-molecule electroluminescence from the Q-band emission of a neutral ZnPc molecule. The NCP emission on the bare surface is also shown as the black curve, highlighting the NCP enhancement effect through the matching of a resonant plasmonic mode with the molecular emission energy. The mechanism of molecular exciton generation in the present system, under a bias of −2.5 V, is believed to be dominated by a charge carrier injection process [4], although the contribution of other mechanisms such as inelastic electron-molecule scattering and plasmon-mediated energy transfer cannot be completely excluded [13].

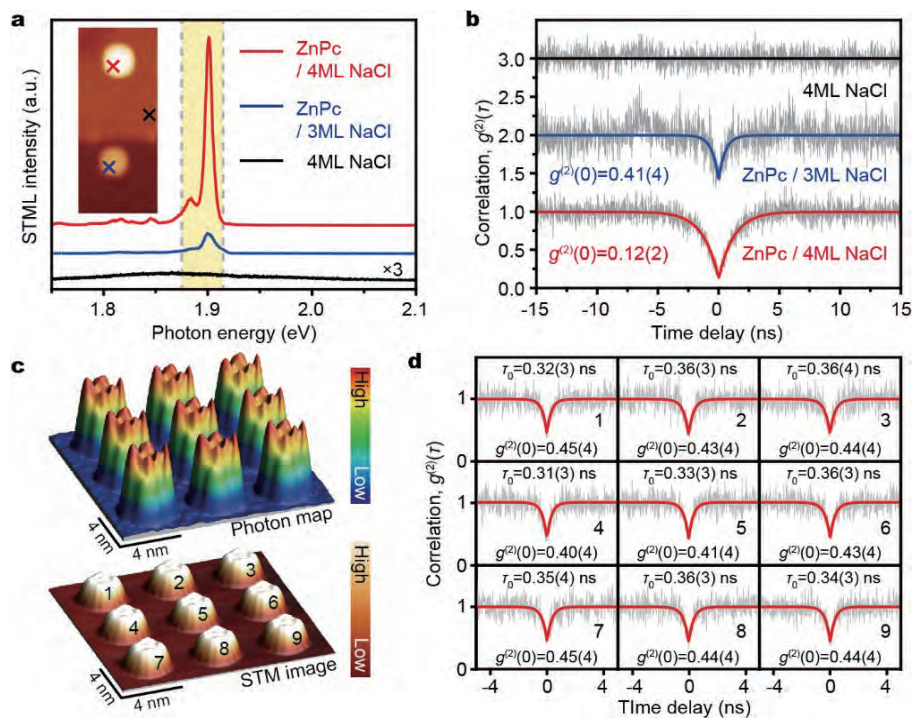

**Fig. 2:** Electrically driven single-molecule single-photon emission. (a) STML spectra of a single ZnPc molecule at different thickness of NaCl layers (−2.5 V, 100 pA, 30 s). The sharp emission peak is attributed to the Q-band emission of neutral ZnPc molecules. (b) Second-order photon correlation measurements of single-molecule electroluminescence. (c) Photon map (top) and STM image (bottom) of a 3 × 3 ZnPc molecular array. (d) Corresponding second-order correlation functions acquired from each individual ZnPc molecule in (c). Adapted with permission from Macmillan Publishers Ltd: [Nature Communications], (L. Zhang, et al., Nature Communications, 8, 580 (2017)) [9].

The single-photon emission properties can be evaluated from the photon correlation measurement with Hanbury Brown and Twiss (HBT) interferometry [21-24]. Figure 2(b) plots the second-order correlation functions [$g^{(2)}(\tau)$] for the three situations shown in Fig. 2(a). An evident antibunching dip can be observed at time zero for the molecular emission when the ZnPc molecule is adsorbed on either 3 or 4 ML NaCl surface, while there is no dip feature in the time histogram of the HBT measurement for the NCP emission on the bare surface. The appearance of a photon antibunching dip indicates clearly the single-photon emission behavior of single-molecule electroluminescence. The purity of single-photon emission can be estimated by fitting the experimental data with an exponential function $g^{(2)}(\tau)=1-[1-g^{(2)}(0)]e^{-|\tau|/\tau_0}$. The value of $g^{(2)}(0)$ is found to be 0.12(2) for ZnPc on 4 ML NaCl and 0.41(4) for ZnPc on 3 ML NaCl, which are both below the threshold value 0.5 for an evident single-photon emission [25, 26]. The time constant ($\tau_0$) of the corresponding fluorescence can also be estimated from the second-order correlation measurements, which is about 1.26(3) ns for ZnPc on 4 ML NaCl and 0.48(5) ns for ZnPc on 3 ML NaCl, respectively. The analysis of these time constants can be used to probe the exciton generation and decay dynamics of a single molecule near metals and inside a plasmonic nanocavity [9, 23].

The advantages of the STM technique in the application of electroluminescence are not only the ability to excite molecular emission by the tunneling electrons and to enhance the signal by the nanocavity, but also include the capability to manipulate single molecules with angstrom precision, allowing us to construct various structures by moving single molecules. As a demonstration, a square array of 3 × 3 ZnPc molecules on 3ML-NaCl with an intermolecular spacing of ∼4.4 nm is constructed and shown in Fig. 2(c). The corresponding photon map measured simultaneously is also shown in the top image of Fig. 2(c), exhibiting similar emission features for different individual molecules. The measured second-order correlation function $g^{(2)}(\tau)$ from each ZnPc molecule in Fig. 2(d) further verifies the identical nature of individual molecular emitters with similar antibunching features and small $g^{(2)}(0)$ values below 0.5. Such a demonstration of identical single-photon emitter arrays illustrates one of the advantages of using molecular single-photon emitters in nanoscale optoelectronic integration.

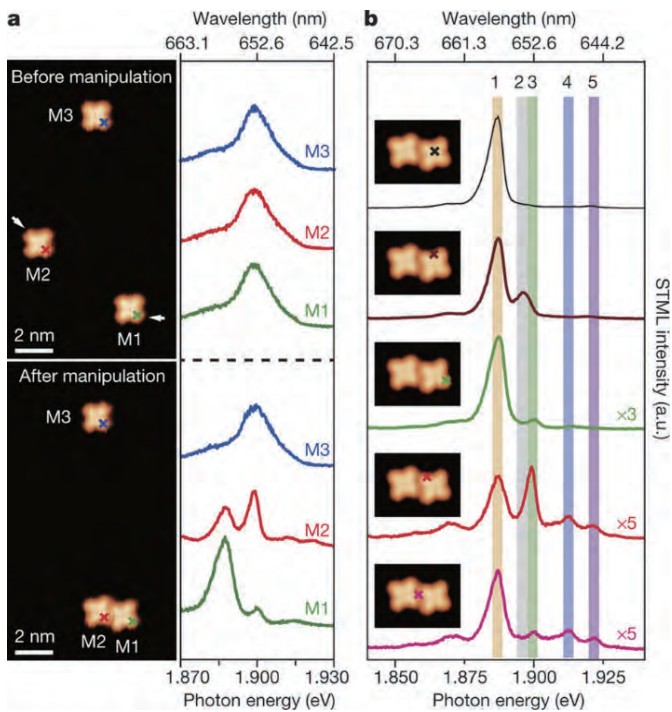

**Fig. 3:** Spectral evolution from a single ZnPc molecule to a dimer. (a) STM images and STML spectra before and after manipulation. (b) Site-dependent STML spectra acquired at different positions within a dimer. Reprinted with permission from Macmillan Publishers Ltd: [Nature], (Y. Zhang, et al., Nature, 531, 623 (2016)) [4].

## VISUALIZING COHERENT INTERMOLECULAR DIPOLE-DIPOLE COUPLING

In the previous sections we have described how to utilize the manipulation capability of STMs to construct a single-photon emitter array, in which each molecule behaves identically to the isolated monomer because the intermolecular distance of about 4 nm is relatively large and there are negligible interactions between individual molecules. Naturally, one could ask what would happen when molecules are brought closer to each other. As shown in Fig. 3(a), when two isolated ZnPc molecules were pushed together to create an artificial molecular dimer with a center-to-center distance of about 1.45 nm, the corresponding STML spectra (right panel in Fig. 3(a)) changed remarkably, and were completely different from that of an isolated monomer. We further checked the spectral features at different representative sites over the ZnPc dimer and found that the emission modes and intensities vary from site to site (Fig. 3(b)). Five major emission peaks can be identified in the STML spectra of a dimer. As shown in Fig. 4(a) and 4(b), in order to give a panoramic view of these five emission modes, we carried out spatially resolved spectroscopic imaging by recording a STML spectrum at each pixel during scanning. Different imaging patterns can be observed for different emission modes, with distinct intensity distributions from the photon image of an isolated monomer. Both the changes in spectral features and the variations in emission patterns indicate that there exists coherent coupling between two single molecules when they form a dimer.

As shown in Fig. 4(c), for a coherently coupled dimer the optical transition energy $\Delta E_{\text{dim}}$ can be written as $\Delta E_{\text{dim}} = \Delta E_{\text{mono}} - \Delta W \pm |J|$, where $\Delta E_{\text{mono}}$ is the transition energy of an isolated monomer, and $\Delta W$ is the energy difference resulting from the van der Waals interaction. The value of $|J|$ represents for the coupling strength, which can be obtained from the point-dipole approximation [27, 28] as $J = [\boldsymbol{\mu}_1 \cdot \boldsymbol{\mu}_2 - 3(\boldsymbol{\mu}_1 \cdot \hat{\mathbf{r}})(\boldsymbol{\mu}_2 \cdot \hat{\mathbf{r}})] / (4\pi\varepsilon_0 r^3)$, where $\boldsymbol{\mu}_1$ ($\boldsymbol{\mu}_2$) is the transition dipole moment of the component monomers (with amplitude $\mu_0$), and $\mathbf{r}$ is the distance vector between two monomer centers with the magnitude $r = |\mathbf{r}|$ and direction $\hat{\mathbf{r}}$. For a single ZnPc molecule it possesses two degenerate excited states and thus two orthogonal transition dipole moments along the $x$ and $y$ axis with the same magnitude. Therefore, the five dipole-coupling modes can be assigned to the following combinations of different dipole moments for a molecular dimer: in-line in-phase (→→), in-line out-of-phase (→←), parallel in-phase (↑↑), parallel out-of-phase (↑↓) and orthogonal (↑→ and →↑). The corresponding coupling strength can be obtained for different dipole phase relations as: $J_{\rightarrow\rightarrow} = -J_{\rightarrow\leftarrow} = -2\mu_0^2/(4\pi\varepsilon_0 r^3)$, $J_{\uparrow\uparrow} = -J_{\uparrow\downarrow} = \mu_0^2/(4\pi\varepsilon_0 r^3)$ and $J_{\uparrow\rightarrow} = J_{\rightarrow\uparrow} = 0$. Figure 4d shows the sequence of the energy levels for different modes, and the emission peaks denoted in Fig. 4a are assigned according to this energy splitting diagram.

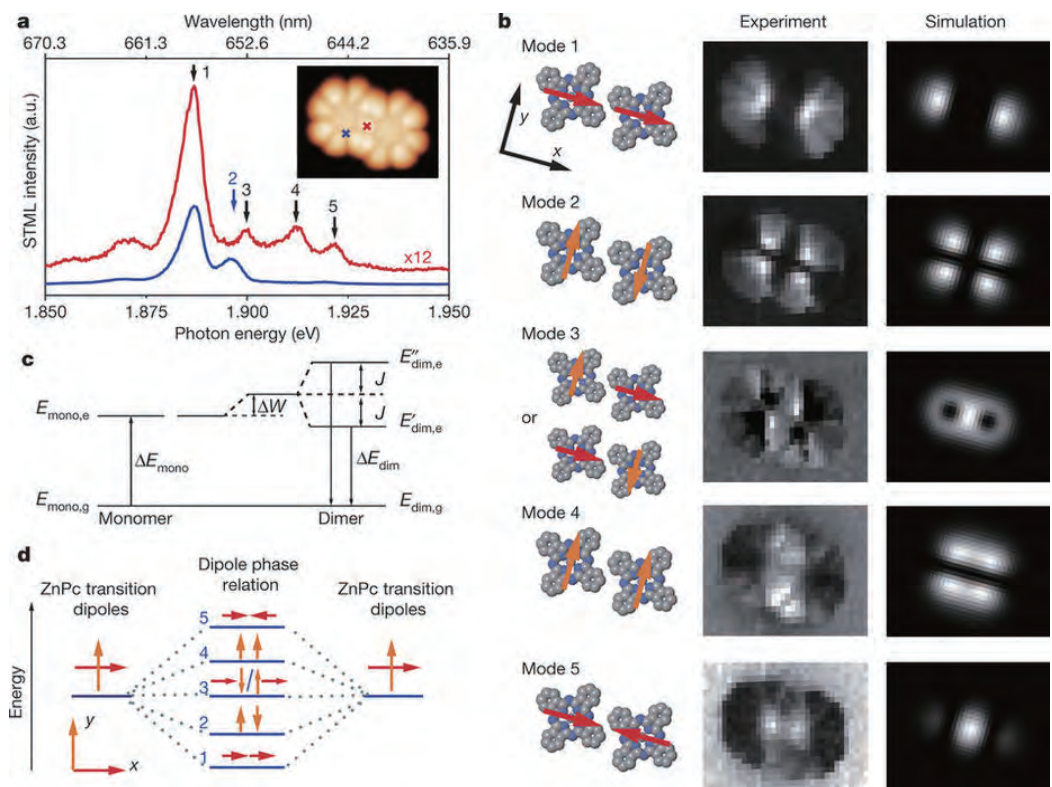

**Fig. 4:** Real-space visualization of coherent dipole-dipole coupling in a ZnPc dimer. (a) STML spectra of a dimer with five labeled emission modes. (b) Photon images (middle) for the five indicated modes, as well as the schematic arrangements of the molecular dipoles (left) and theoretical simulations (right). (c) Exciton band energy diagram of the dimer. (d) Energy splitting diagram for different dipole-dipole coupling modes. Reprinted with permission from Macmillan Publishers Ltd: [Nature], (Y. Zhang, et al., Nature, 531, 623 (2016)) [4].

By comparing the real-space photon images with theoretical simulations based on the emission process (Fig. 4(b)), the main features in the numbers and positions of the emission maxima and nodes can be reproduced by the point-charge modeling [4], while the extra details in the images are likely to arise from the spatial distribution of the electronic densities of the states related to the excitation process. In fact, the in-line and parallel dipole-coupling patterns are in some ways analogous to the $\sigma-$ and $\pi$-type bonding/anti-bonding of molecular orbitals. More specifically, the transition dipole arrangements for mode 1 and mode 4 can be assigned as superradiant modes because of the in-phase coupling, while mode 2 and mode 5 with out-of-phase coupling are subradiant modes that are usually expected to be dark in the far field. Owing to the in-line in-phase coupling configuration and lowest energy level for mode 1, the emission intensity of mode 1 is much stronger than other modes. The possibility of observing the subradiant modes is probably due to the asymmetric image-dipole effect resulted from the tip location. The simultaneous observation of both super-radiant and subradiant emission enables us to evaluate the dipole coupling strength $|J|$ directly, which gives the values of 17.3(3) meV for the in-line and 8.2(3) meV for the parallel configurations, respectively.

The coherent couplings demonstrated above in the spectral and spatial domains imply that the excitation energy, which initially occurred at the monomer directly underneath the STM tip, is rapidly shared between two coupled molecules, oscillating back and forth to yield an entangled dimer. The oscillation frequency $\Omega$ is estimated to be on the order of $10^{13}$ s$^{-1}$, faster than the vibrational relaxation process [29]. In this context, the efficient decoupling of the ZnPc emitters from the metallic substrate through the thin NaCl spacer layer may also facilitate the observation of the coherent dipole-dipole coupling in the molecular dimer. The capability to visualize excitonic coupling in real space with sub-nanometer resolution will open up new opportunities for studying intermolecular interaction and energy transfer at the level of individual molecules.

## SINGLE-MOLECULE FANO RESONANCE IN A PLASMONIC NANOCAVITY

In last two sections, highly localized tunneling electrons are used to excite the molecular emitters and the NCP serves as a local amplifier to enhance the molecular emission signals at the sub-nanometer scale. In fact, the NCP could serve as more than just an amplifier and could also provide valuable information on field-molecule interactions. When a quantum emitter interacts with resonant nanocavity plasmons, the quantum interference resulting from the coherent coupling between the discrete state of the emitter and the continuum-like state of the nanocavity plasmons will lead to changes in the spectral features, which is known as Fano resonance [10, 16-19]. As shown in Fig. 5(a), single-molecule Fano resonance can be achieved by tunneling electron excitations when an STM tip (i.e., the plasmonic nanocavity) is brought close to an isolated ZnPc molecule. As explained above, the single ZnPc molecule is electronically decoupled from the underlying silver substrate by 3 ML NaCl to suppress the fluorescence quenching.

When the STM tip is located above the ZnPc molecule (corresponding to the blue dot in Fig. 5(b) and Situation I in Fig. 5(c)), the molecule is directly excited by tunneling electrons, producing sharp molecular emissions. On the other hand, if the tip is positioned on the bare surface (the green dot in Fig. 5(b) and Situation III in Fig. 5(c)), a broad plasmonic emission is observed. When the tip moves in close proximity to the molecule (the red dot in Fig. 5(b) and Situation II in Fig. 5(c)), a clear dip superimposed over the broad plasmonic spectrum is observed, and the dip position is closely related to the emission peak energy of the ZnPc molecule. This characteristic spectral features of the Fano lineshape reveal the existence of the coherent coupling between the molecular transition and the nanocavity plasmonic resonance.

By exploiting the precise spatial control of the STM technique, the coupling strength between the single molecule and the plasmonic nanocavity can be controlled by tuning the relative distance between them. As an example, when the tip moves away from the edge of a ZnPc molecule along a path schematically shown in Fig. 6(a), the corresponding spectra evolve from a Fano lineshape to a broad plasmon profile (Fig. 6(b)). In other words, the Fano dip gets deeper and wider as the tip moves closer to the molecule, indicating an increase in the coupling strength between the ZnPc molecule and the plasmonic nanocavity. The normalized dip depth is found to decay exponentially with the separation distance (Fig. 6(c)), giving a decay length as small as ∼0.9(1) nm. This suggests that the effective interaction distance of the NCP field is highly confined within ∼1 nm. Through quantitative analysis of the normalized dip depth with a modified dipole coupling model [10], the coupling strength can be estimated, reaching a maximum value as large as ∼15 meV (Fig. 6(d)).

By modifying the STM tip shape, the plasmonic resonant frequency of the nanocavity can be tuned with respect to the molecular transition frequency, which could be further utilized to study the single-molecule Fano resonance for different energy detunings. In Fig. 6(e) the resonant frequency of the NCP is tuned from 620 nm to 690 nm (blue curves), and the corresponding Fano spectra (red curves) exhibit varying asymmetric spectral features. It should be pointed out that even for the zero-detuning condition, the Fano lineshape is still asymmetric, suggesting the plasmon-molecule interaction is beyond the simple dipole-coupling model, involving the influences from the spatial distribution of the molecular transition dipole and higher-order plasmonic modes. Furthermore, an abnormal shift of the Fano dip is observed as large as 3 meV (Fig. 6c), which can be related to the photonic Lamb shift due to the self-interaction of a single molecule in the strong plasmonic field. The ability to spatially control the single-molecule Fano resonance with atomic precision allows to reveal the highly confined nature of the broadband nanocavity plasmon and the coupling strength of the molecule-plasmon interaction, providing new insights into the physical mechanism of the coherent coupling beyond the dipolar coupling model.

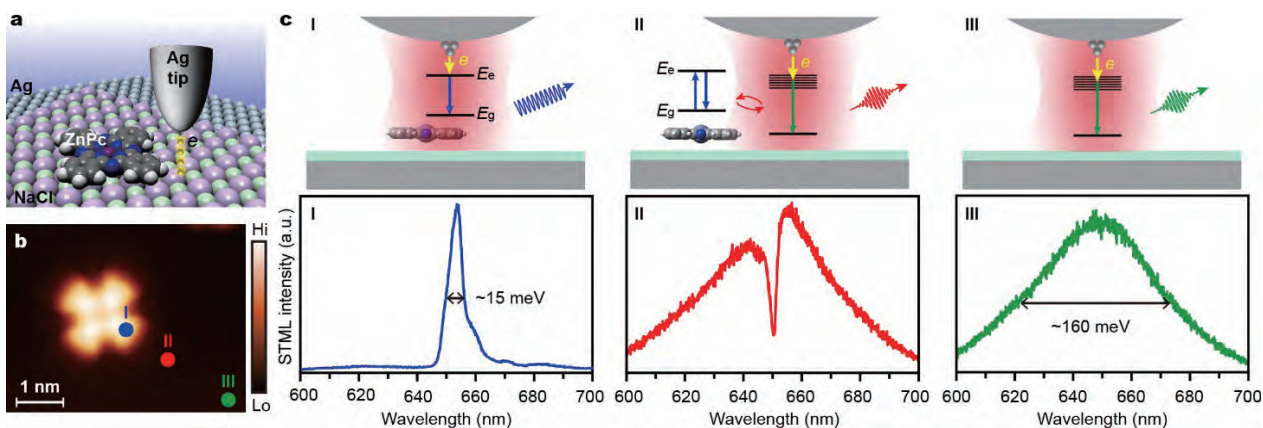

**Fig. 5:** Single-molecule Fano resonance. (a) Schematic experimental setup of achieving single-molecule Fano resonance. (b) STM image of a single ZnPc molecule. (c) Three different configurations (top) and related STML spectra (bottom) corresponding to the dots in (b). Reprinted with permission from Macmillan Publishers Ltd: [Nature Communications], (Y. Zhang, et al., Nature Communications, 8, 15225 (2017)) [10].

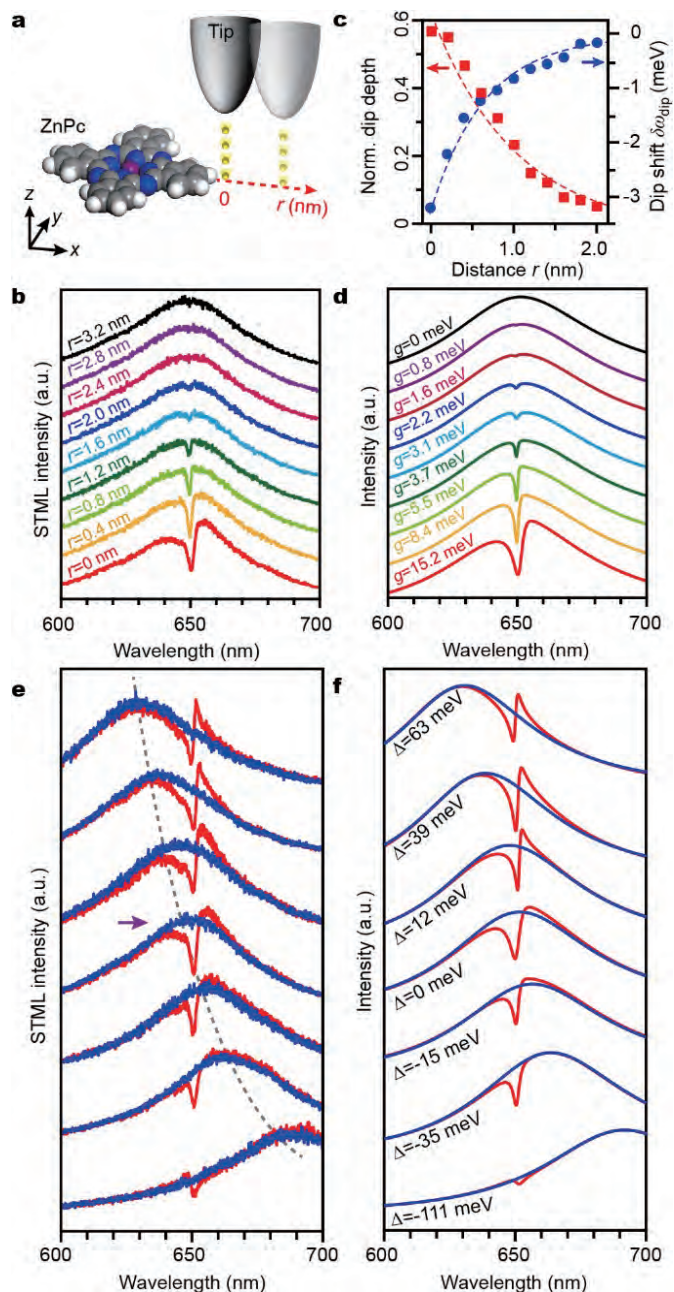

**Fig. 6:** Tunable single-molecule Fano resonance. (a) Schematic of the distance-dependent STML measurements at the edge of the ZnPc molecule. (b) STML spectra along the line trace in (a). (c) Variation of the normalized dip depth and the dip shift with increasing separation distance $r$. (d) Simulated spectra showing the coupling strength. (e) and (f) Experimental and simulated Fano spectra with different energy detunings. Adapted with permission from Macmillan Publishers Ltd: [Nature Communications], (Y. Zhang, et al., Nature Communications, 8, 15225 (2017)) [10].

## CONCLUSION

In conclusion, single-molecule electroluminescence can provide fruitful understandings on the optical properties of single molecules as well as coupled systems, and has potential applications in the fields of point-light sources, single-photon sources, light-harvesting systems and nano-optoelectronics. The realization of STM based single-molecule electroluminescence requires a combined strategy of both efficient electronic decoupling and nanocavity plasmonic enhancement. The observation of photon antibunching effects reveals the single-photon emission nature of single-molecule electroluminescence. Sub-nanometer resolved spectroscopic imaging enables the visualization of coherent intermolecular dipole-dipole coupling in real space. The precise control of single-molecule Fano resonance gives us the ability to extract information about molecule-plasmon interactions. One of the remaining scientific issues to explore in this field is the strong coupling regime between a molecular exciton and the NCP in the STM junction. On the other hand, further combination of STML with ultrafast time-resolved measurements would represent one of the most exciting technical developments, which may help in revealing the dynamics and evolution of the optical processes at the single-molecule level.

**Acknowledgements:** We thank Jian-Guo Hou, Jin-Long Yang, Yi Luo, Bing Wang and Javier Aizpurua for their respective collaborations and discussions. We also thank Li Zhang, Yang Luo, Qiu-Shi Meng and Yun-Jie Yu for performing experiments and analyzing data. This work is supported by the National Basic Research Program of China, the Strategic Priority Research Program of the Chinese Academy of Sciences, the National Natural Science Foundation of China, and the Anhui Initiative in Quantum Information Technologies.

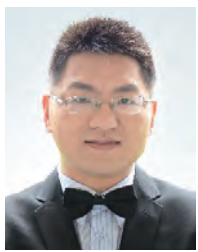

**Yao Zhang** is an associate research fellow at Hefei National Laboratory for Physical Sciences at the Microscale, University of Science and Technology of China. He received his BS from Shanxi University in 2008, his MS from the Institute of Solid Physics, Chinese Academy of Sciences in 2011, and his PhD from the University of Science and Technology of China in 2014. His research interests are in the field of surface plasmon based nano-photonics, focusing on the theoretical calculations of plasmonic properties of metallic nanostructures, the interactions between single molecules and plasmons, and plasmon enhanced spectroscopy.

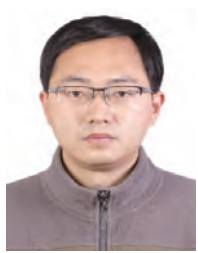

**Yang Zhang** is a professor at Hefei National Laboratory for Physical Sciences at the Microscale, University of Science and Technology of China. He received both his BS and his PhD from the University of Science and Technology of China in 2004 and in 2010, respectively. His research focuses on the investigation of intermolecular interaction and energy transfer with sub-molecular resolved spectroscopic imaging.

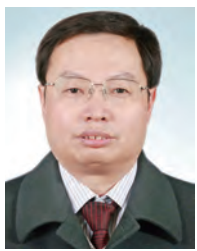

**Zhen Chao Dong** is a full professor at Hefei National Laboratory for Physical Sciences at the Microscale, University of Science and Technology of China (USTC). He received his PhD degree from Fujian Institute on the Structure of Matter, Chinese Academy of Sciences in 1990. After his postdoctoral studies at Iowa State University from 1992 to 1995, he worked at the National Institute for Materials Science (NIMS), Japan from 1996 to 2004 as a senior researcher before joining USTC in 2004. His recent research interests are in the field of single-molecule optoelectronics and nanoplasmonics, particularly on STM based single-molecule electroluminescence and single-molecule Raman scattering as well as energy transfer at the single molecule level, with all achieving spectro-microscopic imaging down to the sub-nanometer scale.